\documentclass[printer]{aa}

\newcommand{\be}{\begin{eqnarray}}
\newcommand{\ee}{\end{eqnarray}}

\usepackage{amssymb}
\usepackage{amsmath}

\usepackage{graphicx}
\usepackage{graphicx}
\usepackage{epsfig}
\usepackage{graphics}

\begin{document}

\title{Gravitational radiation from precessing accretion disks in gamma-ray bursts}
\author{Gustavo E. Romero\inst{1,2,}\thanks{Member of CONICET, Argentina} \and Mat\'{\i}as M. Reynoso\inst{3,\star} \and Hugo R. Christiansen\inst{4}}

\institute{Instituto Argentino de Radioastronom\'{\i}a (IAR), CCT
La Plata  (CONICET), C.C.5, (1894) Villa Elisa, Buenos Aires,
Argentina \and Facultad de Ciencias Astron\'omicas y
Geof\'{\i}sicas, Universidad Nacional de La Plata, Paseo del
Bosque s/n, 1900, La Plata, Argentina \and Instituto de
Investigaciones F\'{\i}sicas de Mar del Plata (CONICET - UNMdP),
Facultad de Ciencias Exactas y Naturales, Universidad Nacional de
Mar del Plata, Dean Funes 3350, (7600) Mar del Plata, Argentina
\and State University of Cear\'a, Physics Dept., Av. Paranjana
1700, 60740-000 Fortaleza - CE, Brazil}

\offprints{G. E. Romero \\ \email{romero@iar-conicet.gov.ar}}

\titlerunning{GW from precessing accretion disks in GRBs}

\authorrunning{G.E. Romero et al.}

\abstract {We study the precession of accretion disks in the
context
     of gamma-ray burst inner engines.}
   {Our aim is to quantitatively estimate the characteristics of gravitational waves produced by the precession of the transient accretion disk in gamma-ray bursts. }
   {We evaluate the possible periods of disk precession caused by the Lense-Thirring effect using an
accretion disk model that allows for neutrino cooling. Assuming
jet ejection perpendicular to the disk plane and a typical
intrinsic time-dependence for the burst, we find gamma-ray light
curves that have a temporal microstructure similar to that observed
in some reported events. The parameters obtained for the
precession are then used to evaluate the production of
gravitational waves.}
{We find that the precession of accretion disks of outer radius
smaller than $10^8$ cm and accretion rates above 1 $M_\odot\; \rm
s^{-1}$ could be detected by Advanced LIGO if they occur at
distances of less than 100 Mpc.}
{We conclude that the precession of a neutrino-cooled accretion
disk in long gamma-ray bursts can be probed by gravitational wave
astronomy. Precession of the disks in short gamma-ray events is
undetectable with the current technology.}


\maketitle

\section{Introduction}
In the central engines of gamma-ray bursts (GRBs), accretion onto a
black hole resulting from the collapse of a massive star (e.g.
Woosly 1993) or a merger of two compact objects (e.g. Mochkovitch
et al. 1993) leads to the formation of a hot and dense, transient accretion disk. This disk can be significantly cooled by
neutrino losses. The accretion of matter, at a rate $\sim
0.1-10$ $M_\odot \; \rm s^{-1}$, is supposed to power the burst, and the
radiative processes in the relativistic jets are expected to
account for the observed light curves. These curves display a wide
variety of time profiles with timescales from milliseconds to
minutes. The usual interpretation of this temporal structure is in
terms of shocks that convert bulk kinetic energy into internal
energy of the particles, which then cool by means of synchrotron and
inverse Compton emission. The shocks can be internal to the jet
and produced by colliding shells with different Lorentz factors
(e.g. Kobayashi, Piran \& Sari 1997; Daigne \& Mochkovitch 1998;
Guetta, Spada \& Waxman 2001) or the result of interactions with
the ambient medium (e.g. Heinz \& Begelman 1999; Fenimore, Madras
\& Nayakshin 1996). Among the observed light curves, however,
there are some that are difficult to explain with the standard
model (e.g. Romero et al. 1999). It has been suggested that the
precession of the jet can play a role in the formation of the
microstructure of both long and short gamma-ray bursts (e.g.
Blackman et al. 1996, Portegies et al. 1999, Fargion 1999, Reynoso
et al. 2006).

Reynoso et al. (2006) developed a model for precessing jets
based on spin-induced precession of the neutrino-cooled massive
disk. The precession of the disk is transmitted to the
relativistic jets, resulting in the peculiar temporal
microstructure of some GRB's light curves.

In this paper, we study an additional effect of this precession of
very massive accretion disks: the production of gravitational
waves. Gravitational wave radiation is expected from gamma-ray
bursts if the gravitational collapse is non-spherical, if there
are strong inhomogeneities in the accretion disk, or, in the case
of short GRBs, as the result of the spiraling and merging of
compact objects (e.g. Mineshinge et al. 2002,  Kobayashi \&
M\'esz\'aros 2003).

The signal we consider here has a different origin from
those previously discussed in the literature, and its specific
features can shed light on the behavior of the innermost regions
of the sources. We demonstrate that gravitational wave astronomy
with instruments such as Advanced LIGO can be used to probe the
Lense-Thirring effect in nearby GRBs.

\section{Accretion disks and spin-spin interaction in GRB engines}

Transient accretion disks are formed in GRB's engines such as
collapsars and mergers of compact objects. The accretion rate in
these disks is expected to vary significantly mostly in the outer
part of the disk, while for the inner disk a constant accretion
rate remains a valid approximation (e.g. Popham 1999, Di Matteo et
al. 2005).

The conservation of mass falling with a velocity $v_r\simeq r
\sqrt{GM_{\rm bh}r^{-3}}$ at a  radius $r$ from the black hole
axis is given by
 \be
 \dot{M}= -2\pi r v_r \Sigma(r),
 \ee
where $\Sigma(r)= 2 \rho(r) H(r)$ is the surface density, $H(r)$
is the disk half-thickness, and $\rho(r)$ is the mass density of
the disk. The conservation of angular momentum and energy can be
used to obtain numerically the functions $\Sigma(r)$ and $H(r)$,
assuming that the heat generated by friction can be balanced by
advection and neutrino emission (Reynoso et al. 2006). If the
orbit of a particle around a spinning black hole is not aligned
with the black hole equator, then the orbit precesses around the
spinning axis. This is called the Lense-Thirring effect (Lense \&
Thirring 1918), and originates from the dragging experienced by the
inertial frames close to the rotating black hole. In accretion
disks, the action of viscous torques leads to the alignment of the
very inner part of the disk with the black hole equator
plane (Bardeen \& Petterson 1975).

If the transient accretion disk in GRBs is formed misaligned with
respect to the equator of a rapidly spinning black hole, then the
accretion disk will develop precession. It has been argued that
the disk precesses approximately like a rigid body, i.e., it does
not present warping, when the disk Mach number is $\mathcal{M}<5$
(Nelson \& Papaloizou 2000). This condition is fulfilled in the
accretion disks of GRBs.  The precession of the disk leads to the
precession of the jets, yielding a source of temporal variability
and microstructure in the signal (Reynoso et al. 2006).

Neglecting any nutation movement, the precession period of the
disk $\tau_\mathrm{p}$ can be related to its surface density as (see Liu \& Melia 2002, Caproni et al. 2004, Reynoso et al. 2006):
 \be
  \tau_p=\int_{0}^{2\pi} \frac{L_{\rm d}}{T_{\rm d}}\sin \theta d\phi = 2
  \pi \sin \theta \frac{L_{\rm d}}{T_{\rm d}},
 \ee
where the magnitudes of the disk angular momentum $L_{\rm d}$ and the precessional torque $T_{\rm d}$ applied to the disk are
 \be L_{\rm d} &=& 2
 \pi\int_{R_\mathrm{ms}}^{R_\mathrm{out}} \Sigma(r) \Omega_\mathrm{k}(r) r^3 \ dr, \\
 T_{\rm d}&=&4\pi^2 \sin{\theta}
 \int_{R_\mathrm{ms}}^{R_\mathrm{out}} \Sigma(r) \Omega_\mathrm{k}(r) \nu_{p,\theta}(r)
 r^3 \ dr.
 \ee
Here, if $a_*$ is the spin parameter,
  \be \Omega_\mathrm{k}(r)=\frac{c^3}{GM_\mathrm{bh}} \left[
\left( \frac{r}{R_\mathrm{g}} \right)^{3/2}+a_*  \right]^{-1} \ee
is the relativistic Keplerian angular velocity,
$R_\mathrm{g}=GM_\mathrm{bh}/c^2$ is the gravitational radius, and
\be \nu_{p,\theta}=\frac{\Omega_k(R)}{2\pi} \left[1 - \sqrt{1\mp
4a_*\left(\frac{R_\mathrm{g}}{r}\right)^{1/2}+3a_*^2
\left(\frac{R_{\rm g}}{r}\right)^2}\right] \ee is the nodal
frequency obtained by perturbing a circular orbit in the Kerr
metric (Kato 1990). The precessing part of the disk ends at an
outer radius $R_{\mathrm{out}}$, extending from an inner radius
$R_\mathrm{ms}= \xi_{\rm ms}R_\mathrm{g}$, where \be
\xi_\mathrm{ms}=3 +A_2 \mp [(3-A_1)(3+A_1+2A_2)]^{1/2}, \ee with
\be A_1=1+(1-a_*^2)^{1/3}[(1+a_*)^{1/3}+(1-a_*)^{1/3}], \ee and
\be A_2=(3a_*^2+A_1^2)^{1/2}. \ee The minus sign in the expression for $\xi_{\rm ms}$
corresponds to prograde motion ($a_*>0$), whereas the plus sign
corresponds to retrograde motion ($a_*<0$).

\begin{figure}[h]
  \includegraphics[trim=15 105 0 22,clip, width=0.68\textwidth]{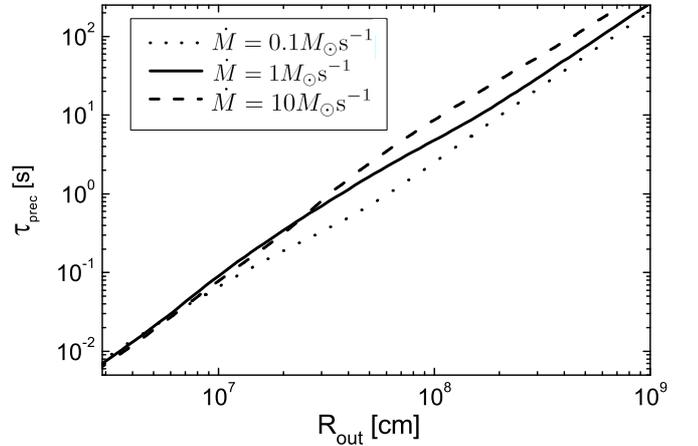} 
  \caption{Precession period as a function of the outer radius of the accretion disk for $\dot{M}=0.1 M_\odot \; {\rm s}^{-1}$ (dotted line), $\dot{M}=1 M_\odot \; {\rm s}^{-1}$ (solid line), and $\dot{M}=10 M_\odot \; {\rm s}^{-1}$ (dashed line). }
  \label{Fig_taup}
\end{figure}

In Fig. \ref{Fig_taup}, we show the precession period obtained as a
function of the disk outer radius for different accretion rates.

\section{Gravitational waves from precessing disks in GRBs}

{We attempted to characterize the emission of gravitational waves (GWs) from precessing accretion disks in GRB engines. 
Precession leads to observable features in the gamma-ray light curves. Its effects are illustrated in the following subsection, and the necessary formulae for GW emission are presented in Sect. 3.2.}

\subsection{Specific models of precessing accretion disk in GRBs}

{We assumed that the precession of accretion disks in GRBs is transmitted to the jets.} 
The electromagnetic emission is generated in the jet by
synchrotron and inverse Compton processes. The dependence of the
gamma-ray luminosity on the angle with respect to the jet axis $\psi$
is described by the expression (Portegies-Swart et al.
1999)

 \begin{multline}
 L(\psi)= \frac{27}{4}\left[e^{-0.6 \Psi(x)}-\frac{8}{9}\right]  \\ \times \left[ e^{-0.3 \Psi(x)}- e^{-6.3 \Psi(x)}\right]\frac{d\Psi(x)}{dx}, \label{Lum}
 \end{multline}
where $x=10 \sin\psi$ and $\Psi(x)= \frac{1}{6}\ln(1+4x^2)$. The
intrinsic temporal dependence of the signal is characterized by a FRED (Fast Rise and Exponential Decay) behavior,
 \be
   I(t) = N_I \left( 1-e^{-\frac{t}{\tau_{\rm rise}}} \right) \left\{ \frac{\pi}{2}-
   \tan^{-1} \left[ \frac{t-\tau_{\rm plat}}{\tau_{\rm dec}} \right]\right\}, \label{Idt}
 \ee
where $N_I$ is a normalization constant such that the maximum of
the signal corresponds to unity, and $\tau_{\rm rise}$, $\tau_{\rm
plat}$, and $\tau_{\rm dec}$ are the timescales for the initial
rise, plateau, and decay, respectively.

We considered two specific
GRBs as examples: the short burst GRB 990720 and the long burst GRB 990712. We
found for these bursts, a proper set of the relevant timescales and precession period. Then, using the model of Reynoso et al.
(2006) we reproduced the observed light curves.

\begin{table}[h]
\caption{Parameters used for the example GRB events.} 
\centering 
\begin{tabular}{c c  c} 
\hline\hline 
Parameter &  GRB 990712    &   GRB 990720  \\ [0.5ex] 
\hline 
$\tau_{\rm prec}[{\rm s}]$         & 1.0   & 0.2  \\
$\tau_{\rm rise}[{\rm s}]$   & $2.0$ & $0.04$  \\
$\tau_{\rm plat}[{\rm s}]$   & $10.0$ & $0.5$  \\
$\tau_{\rm dec}[{\rm s}]$   & $4.8$ & $0.14$  \\
$\dot{M}[M_\odot{\rm s}^{-1}]$   & $ 1-10$ & $ 0.1$  \\
$R_{\rm out}[10^7{\rm cm}]$   & $3.3-3.7$ & $ 2.1$  \\ [1ex]
\hline\hline 
\end{tabular}
\label{table:nonlin} 
\end{table}

\begin{figure}[h]
  \includegraphics[trim=18 0 0 0,clip, width=0.55\textwidth]{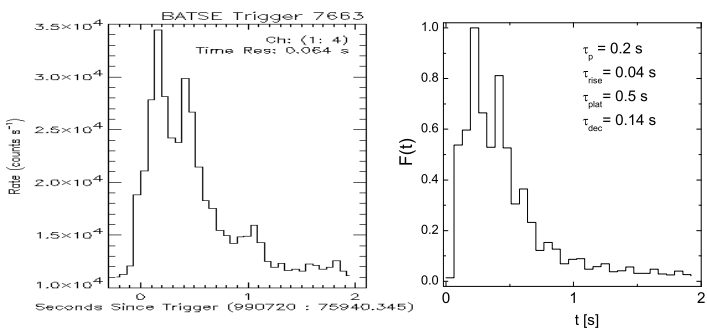} 
  \caption{GRB 990720 light curve (left panel) and light curve obtained with a precessing disk (right panel). }
  \label{FigGRBshort}
\end{figure}

\begin{figure}[h]
  \centering
  \includegraphics[trim=17 0 0 0,clip, width=0.55\textwidth]{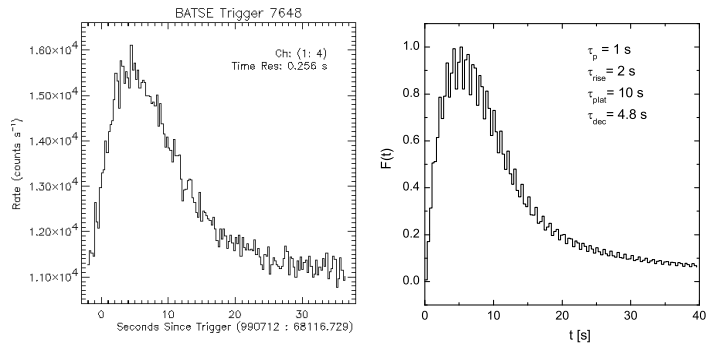} 
  \caption{GRB 990712 light curve (left panel) and light curve obtained with a precessing disk (right panel). }
  \label{FigGRBlong}
\end{figure}

The original time profiles corresponding to the mentioned events
and the light curves obtained using
$F(t)=I(t)L(\psi(t))$ where the effect of precession is given by $\phi(t)= 2\pi (t/\tau_{\rm prec})$, are shown in Figs.
\ref{FigGRBshort} and  \ref{FigGRBlong}. In both cases, the
observer is located at $\theta_{\rm obs}=2^\circ$ with respect to the
$z$-axis perpendicular to the black hole equator, and the angle
$\psi(t)$ between the jet and the observer is time-dependent
because of the precession.

\subsection{General formulae for gravitational wave emission}

An axissymmetric body (i.e., with inertial moments $I_1=I_2$) in
precession emits gravitational waves with an amplitude given by
(Zimmermann \& Szedenits 1979, Maggiore 2008)
 \be
 h_{\rm prec}(t)= h_+(t)+h_\times(t), \label{hprec}
 \ee
where
 \be
 h_{+}(t)= F_{+,1} \cos{\Omega t} +F_{+,2} \cos{2\Omega t},  \\
 h_{\times}(t)= F_{\times,1} \sin{\Omega t} +F_{\times,2} \sin{2\Omega t},
 \ee
with
 \be
  F_{+,1}&=& h'_0 \sin{2\alpha} \sin{\iota} \cos{\iota}\\
  F_{+,2}&=& 2 h'_0 \sin^2{\alpha} (1+\cos^2{\iota}) \\
  F_{\times,1}&=& h'_0 \sin{2\alpha} \sin{\iota} \\
  F_{\times,2}&=& 4 h'_0 \sin^2{\alpha} \cos{\iota},
 \ee
and
 \be
 h'_0= -\frac{G}{c^4}\frac{(I_3-I_1)\Omega^2}{d}.\label{h0}
 \ee
Here, $\alpha$ is the angle between the angular momentum of the
disk and that of the black hole, $\iota$ is the angle between the
$z$-axis of the detector and the signal direction of arrival, and
$d$ is the distance to the radiating body. The principal inertia
moments are
 \be
 I_3&=& \int_V (x^2+y^2) \rho(r)\\
 I_1&=& \int_V (z^2+y^2) \rho(r).
 \ee
The frequency of the gravitational waves are $f_1=\Omega/(2\pi)$ and
$f_2=2\Omega/(2\pi)$, which are related to the angular momentum of the body
by
 \be
 \Omega= \frac{L}{I_1}.
 \ee

\begin{figure}[h]
  \centering
  \includegraphics[trim=30 0 0 0,clip, width=0.5\textwidth]{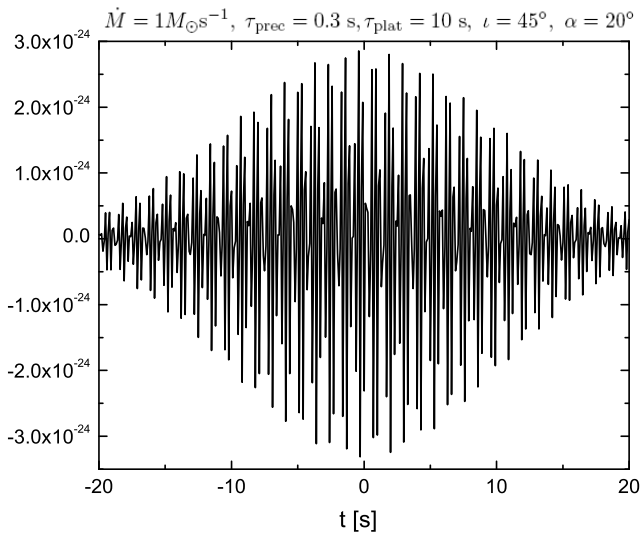} 
  \caption{A typical GW waveform that can be produced by a precessing GRB accretion disk. The accretion rate is $\dot{M}=1 M_\odot {\rm s}^{-1}$, the precession period is $\tau_{\rm prec}=0.3$ s, and the duration of the event is given by the timescale $\tau_{\rm plat}=10$ s. }
  \label{hdtfig}
\end{figure}

{Since we are dealing with bursting events, the GW signal is expected to be significant for a time $\tau_{\rm plat}$.
Therefore, we modulated the signal of Eq. (\ref{hprec}) using a Gaussian}
 \be
 h(t)=h_{\rm prec}(t) \ e^{-\frac{t^2}{2\tau_{\rm plat}^2}}.\label{hdet}
 \ee 
{This is usually adopted to describe the GW signal from bursting sources (e.g. Acernese 2008, Maggiore 2004).
The angular frequencies that contribute to the waveform of Eq. (\ref{hdet}) can be obtained from its Fourier transform 
$\tilde{h}(\omega)= \tilde{h}_+(\omega) + \tilde{h}_\times(\omega) $, where}
 \begin{multline}
 \tilde{h}_+(\omega)= \frac{\tau_{\rm plat}}{2} \left[ F_{+,1}\left(e^{ \frac{-\tau^2_{\rm plat}}{2}(\omega+\Omega)^2 }+ e^{\frac{-\tau^2_{\rm plat}}{2}(\omega-\Omega)^2}\right) \right. \\
 + \left.F_{+,2}\left(e^{\frac{-\tau^2_{\rm plat}}{2}(\omega+2\Omega)^2 }+ e^{ \frac{-\tau^2_{\rm plat}}{2}(\omega-2\Omega)^2 }\right) \right]
 \end{multline}
 and
 \begin{multline}
 \tilde{h}_\times(\omega)= \frac{i \tau_{\rm plat}}{2} \left[ F_{\times,1}\left(e^{ \frac{-\tau^2_{\rm plat}}{2}(\omega+\Omega)^2 }+ e^{\frac{-\tau^2_{\rm plat}}{2}(\omega-\Omega)^2}\right) \right. \\
 + \left.F_{\times,2}\left(e^{\frac{-\tau^2_{\rm plat}}{2}(\omega+2\Omega)^2 }+ e^{ \frac{-\tau^2_{\rm plat}}{2}(\omega-2\Omega)^2 }\right) \right].
  \end{multline}
{It can be seen from these expressions that the ranges of frequencies that contribute to the signal are centered around $\Omega$ and $2\Omega$,
and the width of the ranges is $\tau_{\rm plat}^{-1}$. Given the typical durations of GRBs, the frequency spread is narrow for all bursts except for those with durations much shorter than $1$ s.
In Fig. \ref{hdtfig} we show an example waveform obtained with Eq. (\ref{hdet}) for an accretion rate $\dot{M}=1 M_\odot {\rm s}^{-1}$, a precession period $\tau_{\rm prec}=0.3$ s, $\tau_{\rm plat}= 10$ s, $\iota=45^\circ$, $\alpha=20^\circ$, and $d=100$ Mpc.}

\section{Detectability}

\begin{figure*}[]
  \centering
  \includegraphics[trim=15 200 0 18,clip, width=0.99\textwidth, keepaspectratio]{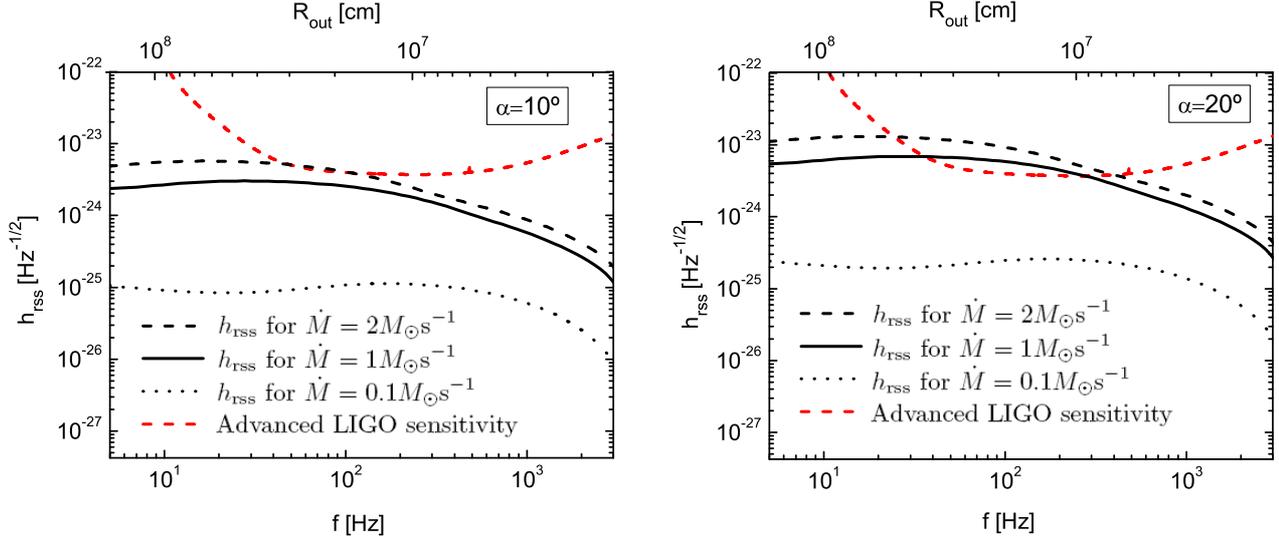} 
  \caption{Gravitational wave rss amplitude for different accretion rates and Advanced LIGO sensitivity (red dashed line) as a function of the gravitational wave frequency for $\alpha=10^\circ$ (left panel) and for $\alpha=20^\circ$ (right panel). The corresponding outer radius of the accretion disk is indicated in the upper horizontal axis.}
  \label{Fig_h0deRout}
\end{figure*}

We now consider the gravitational waves produced by the disk
precession in events with the characteristics of those discussed
in the previous section. In Table 1, we list the main parameters
for the short and long GRBs taken as examples. We estimated the
amplitude of the gravitational waves using Eq. (\ref{h0}),
considering different accretion rates. 
{In order to assess the detectability, we calculated the root-sum-square amplitude (e.g. Acernese 2008, Maggiore 2004)}
 \be
 h_{\rm rss}(f)= \sqrt{\int_{-\infty}^{\infty}dt( h_+^2+ h_\times^2}),
 \ee
{where $f=\Omega/(2\pi)$. For illustration, we choose the value $\iota=45^\circ$ for the angle between the line of sight 
and the z-axis of the detector, and we take $\alpha=10^\circ-20^\circ$ to be the angle between the angular momentum of the disk and that of the black hole.} {In Fig. \ref{Fig_h0deRout}
we plot the root-sum-square amplitude $h_{\rm rss}$,
as a function of one of the allowed frequencies, $f=\Omega/(2\pi)$}, and
also as a function of the outer radius of the inner precessing
disk. In the figure, we include the expected sensitivity for
Advanced LIGO (Shoemaker 2010). 

The parameters used to find $\Sigma(r)$ and
$H(r)$ are $M_{\rm bh}= 3 M_\odot$, $a_*=0.1$, a viscosity
parameter $\alpha=0.1$, and the different mass-loss rates $\dot{M}=\{0.1 M_\odot \;{\rm s}^{-1}, 1 M_\odot \; {\rm
s}^{-1},10 M_\odot \; {\rm s}^{-1}\}$ . The distance to the GRB is
taken to be $d= 100\; {\rm Mpc}$.

As can be seen from Fig. \ref{Fig_h0deRout}, there are higher probabilities of detection for accretion rates higher than $1M_\odot \;
{\rm s}^{-1}$ and outer radii between $10^{7}$ and $10^{8}$ cm.
When the accretion rate is very high, the disk may become advection
dominated rather than cooled by neutrino emission (Liu et al.
2008). The dynamics in the gravitational field, however, is not
affected.  High accretion rates can be sustained only in long
GRBs, so we conclude that there is only a good prospect for
detection of gravitational waves from precessing disks of nearby
($d<100$ Mpc) and long events. These events are likely to be related to
the death of very massive stars, so the host galaxies should have
active star-forming regions.

Two low-luminosity long GRBs (980425 and 060218) were already
observed at distances of $\sim$ 40 Mpc and $\sim$ 130 Mpc,
respectively (Corsi \& M\'esz\'aros 2009). The local rate of
long GRBs is estimated to be $\sim$ 200 Gpc$^{-3}$ yr$^{-1}$ (e.g.
Liang et al. 2007, Virgili et al. 2009). INTEGRAL has detected a
large proportion of faint GRBs that have been inferred to be local (Foley et al.
2008). All this suggests that the detection of precessing disks of
GRBs through their gravitational emission is possible in the near
future.

\section{Summary and conclusions}

We have studied long and short gamma-ray bursts showing that the
presence of microvariability in their light curves can be caused by
the precession of the transient accretion disk. 
We then considered precessing disks of GRBs as a source of GWs and estimated the
resulting waveform produced by the phenomenon.

Our results indicate that if the outer radius of the
precessing disk is between $10^{7}$ and $10^{8}$ cm and the
accretion rate higher than $1\; M_\odot \; {\rm s}^{-1}$, the
gravitational waves can be detected from distances of 100 Mpc or
less. We conclude that only in relatively nearby and long GRBs can the
precession of the accretion disk be studied by gravitational
wave astronomy. The detection of one event of this class can be
used to test the Lense-Thirring effect in the strong field limit.

\begin{acknowledgements}
We thank O.A. Sampayo for useful comments on GW physics.
This research was supported by the Argentine Agencies CONICET and
ANPCyT through grants PIP 112-200901-00078, PIP 112-200801-00587,
and PICT-2007-00848 BID 1728/OC-AR. GER acknowledges additional 
support from the Ministerio de Educaci\'on y Ciencia (Spain) under grant AYA 2007-68034-C03-01, FEDER funds,
and HRC acknowledges financial support of FUNCAP, Brazil.

\end{acknowledgements}

\end{document}